\documentclass[aps,prd,twoside,twocolumn,nofootinbib,%
showpacs,floatfix]{revtex4-1}

\usepackage{amsmath,amssymb}
\usepackage{graphicx,bm}
\usepackage{slashed,color}

\begin{document}

\title{\boldmath Axial anomaly and the $\delta_{LT}^{}$ puzzle}

\author{Nikolai Kochelev}%
\email{kochelev@theor.jinr.ru}

\affiliation{Bogoliubov Laboratory of Theoretical Physics,
Joint Institute for Nuclear Research, Dubna, Moscow region, 141980
Russia}

\author{Yongseok Oh}%
\email{yohphy@knu.ac.kr}

\affiliation{Department of Physics,
Kyungpook National University, Daegu 702-701, Korea}

\date{}

\begin{abstract}
The axial anomaly contribution to generalized longitudinal-transverse
polarizability $\delta_{LT}^{}$ is calculated within Regge approach.
It is shown that the contribution from the exchange of the $a_1^{}(1260)$
Regge trajectory is nontrivial and might have the key role to
explain the large difference between the predictions of chiral
perturbation theory and the experimental data for the
neutron $\delta_{LT}^{}$.
We also present the prediction for the proton $\delta_{LT}^{}$ that will
be measured at the Thomas Jefferson National Accelerator Facility in near future.
\end{abstract}

\maketitle

The measurement of spin-dependent lepton-nucleon cross sections
provides very important information about nucleon structure and is
a very useful tool for examining the validity of various
approaches for description of the QCD effects at large distances
between quarks and gluons. One of such successful approaches is
chiral perturbation theory ($\chi$PT) that enjoys successful
description of hadron physics at small momentum transfer and at
low energy. However, it was found that the recent
predictions~\cite{KSV02,BHM03} of various versions of $\chi$PT
show a strong deviation from the data for generalized
longitudinal-transverse polarizability ($\delta_{LT}$) of the
neutron at low $Q^2$ measured by the E94010 Collaboration of the
Thomas Jefferson National Accelerator Facility
(TJNAF)~\cite{JLE94010-04}. The generalized
longitudinal-transverse polarizability can be evaluated from a
combination of the spin-dependent structure functions $g_1^{}$ and
$g_2^{}$, and is an ideal quantity to test models for hadron
reactions at low $Q^2$. Therefore, such a large deviation is a
serious challenge to the $\chi$PT approach to low energy
reactions, and finding the possible sources for this discrepancy
is an important open question.

In this paper, we suggest a possible way to resolve this puzzle based on the
consideration of the axial anomaly contribution to $\delta_{LT}^{}$ arising
through the $t$-channel $a_1^{}$ Regge-trajectory exchange to the
spin-dependent forward scattering amplitude for $\gamma^*N \to \gamma^*N$.
We will show that
this contribution is nontrivial and has a crucial role to bring a
$\chi$PT prediction for the neutron to the measured data.
For further testing of this idea, we also present the prediction for the proton
$\delta_{LT}^{}$ in this approach, which can be examined by the
planned experiment at TJNAF~\cite{Slifer09}.

\begin{figure}[t]\centering
\vskip -0.4cm
\includegraphics[width=0.45\textwidth,angle=0,clip]{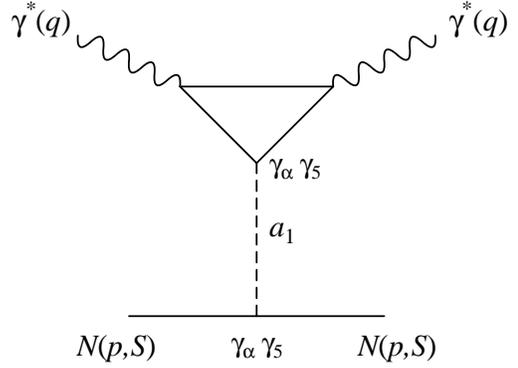}
\caption{\label{fig:diagram}
The $a_1^{}$-exchange contribution to the virtual
photon-nucleon forward scattering amplitude.}
\end{figure}

It is widely known that the axial anomaly~\cite{Adler69,BJ69} plays a very
important role in hadron physics.
Originated from the quark triangle diagram, the axial anomaly determines
the $\pi^0 \to \gamma\gamma $ decay width and provides a crucial role to the
``spin crisis." (For a review, see, for example, Ref.~\cite{AEL95}.)
In the present paper, we calculate its contribution to
$\delta_{LT}^{}$, which is induced by the $t$-channel exchange of
the $a_1(1260)$-meson Regge trajectory as depicted by the diagram
in Fig.~\ref{fig:diagram}.

In the hadronic tensor of two electromagnetic currents,
\begin{equation}
W_{\mu\nu}=\frac{1}{4\pi}\int d^4x\, e^{ipx} \langle p\, S \mid \left[J_\mu(x),J_\nu(0)
\right] \mid p\, S \rangle,
\end{equation}
the spin-dependent part is
\begin{eqnarray}
W^{\rm spin}_{\mu\nu} &=&
i\epsilon_{\mu\nu\alpha\beta}\frac{q^\alpha S^\beta}{(p\cdot q)}\,g_1^{}(x,Q^2)
\nonumber \\ && \mbox{}
+ i\epsilon_{\mu\nu\alpha\beta}
\frac{q^\alpha \left[ (p\cdot q) S^\beta-(S\cdot q)p^\beta \right]}{(p\cdot q)^2}
\,g_2^{}(x,Q^2),
\nonumber \\ \label{eq:spin}
\end{eqnarray}
where $q$ and $p$ are the momenta of the virtual photon and the nucleon, respectively.
In Eq.~(\ref{eq:spin}), $x=Q^2/(2p\cdot q)$ with $Q^2=-q^2$,
and $g_1^{}(x,Q^2)$ and $g_2^{}(x,Q^2)$ are spin-dependent nucleon structure functions.
The spin vector of the nucleon $S_\mu$ is normalized as $S^2=-M^2_N$ with
$M_N$ being the nucleon mass.
The details can be found, for example, in Ref.~\cite{BK96d}.
In terms of these structure functions the generalized longitudinal-transverse
polarizability $\delta_{LT}^{}$ can be written in the following form:
\begin{equation}
\delta_{LT}^{}(Q^2) =
\frac{16\alpha M_N^2}{Q^6} \int_0^{x_0^{}} dx\, x^2 \left[ g_1^{}(x,Q^2)
+ g_2^{}(x,Q^2) \right ],
\label{def}
\end{equation}
where $x_0^{} = Q^2/(2M_N\nu_0^{})$, and $\nu_0^{} = m_\pi +
(m^2_\pi+Q^2)/(2M_N) $ is the threshold photon energy for
$\pi$-meson production. Here, $\alpha = e^2/4\pi$ is the
electromagnetic fine structure constant.

The spin-dependent part of the forward Compton amplitude is
\begin{eqnarray}
T^{\rm spin}_{\mu\nu} &=&
i\epsilon_{\mu\nu\alpha\beta} \frac{q^\alpha S^\beta}{M^2_N} A_1(Q^2,\nu)
\nonumber \\ && \mbox{}
+ i\epsilon_{\mu\nu\alpha\beta}
\frac{q^\alpha \left[ (p\cdot q) S^\beta-(S\cdot q)p^\beta \right]}{M^4_N}
A_2(Q^2,\nu),
\nonumber \\
\label{compton}
\end{eqnarray}
which is related to the structure functions $g_1^{}$ and $g_2^{}$ as
\begin{eqnarray}
g_1^{}(x,Q^2) &=& \frac{(p\cdot q)}{2\pi M^2_N}
\,\mbox{Im} \left[A_1(Q^2,\nu)\right],
\nonumber \\
g_2^{}(x,Q^2) &=& \frac{(p\cdot q)^2}{2\pi M^4_N}
\,\mbox{Im} \left[A_2(Q^2,\nu)\right],
\label{str}
\end{eqnarray}
where $\nu$ is the photon energy.
Since we are interested in the behavior of $\delta_{LT}^{}$ at low $Q^2$,
it is rather convenient to use the variable $\nu$ instead of the
variable $x$ defined for deep inelastic scattering.
Then, the generalized longitudinal-transverse
polarizability $\delta_{LT}^{}$ of Eq.~(\ref{def}) can be written as
\begin{equation}
\delta_{LT}^{}(Q^2) =
\frac{2\alpha}{M_N} \int_{\nu_0^{}}^{\infty} \frac{d\nu}{\nu^4}
\left[g_1^{}(\nu,Q^2) + g_2^{}(\nu,Q^2) \right].
\label{def1}
\end{equation}

The contribution of the diagram in Fig.~\ref{fig:diagram} to the spin-dependent
forward Compton scattering amplitude is
\begin{eqnarray}
T^{{\rm spin},a_1^{}}_{\mu\nu} &=& g_{a_1^{}NN}^{}
\bar N(p,S) \gamma_\beta^{} \gamma_5^{} N(p,S)
\nonumber \\ && \mbox{} \times P_{a_1^{}}^{\alpha\beta}(t=0,\nu) R_{\mu\nu\alpha}(Q^2),
\label{cont}
\end{eqnarray}
where $g_{a_1^{}NN}^{}$ denotes the coupling constant of the $a_1^{}$-meson
coupling to the nucleon,
$P_{a_1}^{\alpha\beta}(t,\nu)$ is the $a_1^{}$-meson propagator, and
$R_{\mu\nu\alpha}(Q^2)$ is the triangle part of the diagram in
Fig.~\ref{fig:diagram}, which is related to the axial anomaly.
For the point-like $a_1^{}$-quark vertex in the triangle graph%
\footnote{For large values of $Q^2\geq 1$~GeV one should take into account
the non-locality of this vertex, which will lead to the suppression of
the form factor at large $Q^2$. There can be another form for the $a_1^{}$-quark
vertex as suggested in Ref.~\cite{BKRS99}, which, however, does not contribute 
to the forward Compton scattering amplitude.}
we use the well-known formula~\cite{AEL95},
\begin{equation}
R^{\mu\nu\alpha}(Q^2)=i\frac{\mathcal{A}}{2\pi^2} g_{a_1^{}qq}^{}
q_\tau^{} \epsilon^{\mu\nu\tau\alpha} F(Q^2,m_q^2),
\label{top}
\end{equation}
where $\mathcal{A} = N_c (e^2_u-e^2_d)/\sqrt{2}$.
Here, $N_c$ is the number of color, $e_q$ is the electric charge of $q$-quark,
$g_{a_1^{} qq}^{}$ is the coupling constant of the $a_1^{}$ meson and the quark,
and $m_q^{}$ is the constituent quark mass in the triangle diagram.
In Eq.~(\ref{top}), the form factor $F(Q^2,m_q^2)$ is
\begin{equation}
F(Q^2,m_q^2) = 1 + \frac{2m_q^2}{Q^2\rho(Q^2)}
\log \left( \frac{\rho(Q^2)-1}{\rho(Q^2)+1} \right),
\label{form}
\end{equation}
where
\begin{equation}
\rho(Q^2)=\sqrt{1+\frac{4m_q^2}{Q^2}}.
\end{equation}
Note that this form factor vanishes in the real photon limit, i.e.,
$F(Q^2,m_q^2) \to 0$ as $Q^2 \to 0$.
This feature is in agreement with the Landau-Yang theorem~\cite{Landau48,Yang50}
that prohibits the axial-vector meson decay into two real photons.
Therefore, the $a_1^{}$-meson exchange does not contribute to the real photon
forward Compton scattering amplitude.
As a result, the $a_1^{}$ exchange does not alter, for example, the
Gerasimov-Drell-Hearn sum rule and $\delta_{LT}^{}$ at $Q^2=0$.

At high center-of-mass photon-nucleon energy, the $a_1$-meson propagator
should be replaced by its Regge propagator as~\cite{GLV97}
\begin{equation}
\frac{g^{\alpha\beta}-k^\alpha k^\beta/M^2_{A}}{t-M^2_A}
\Rightarrow P_{A^{}}^{\alpha\beta}(t,\nu)_{\rm Regge}
\end{equation}
where $A$ stands for the $a_1^{}$ meson, and
\begin{eqnarray}
P_{A^{}}^{\alpha\beta}(t,\nu)_{\rm Regge}
&=& \left( g^{\alpha\beta}-\frac{k^\alpha k^\beta}{M^2_{A}} \right)
\left( \frac{s}{s_0^{}} \right)^{\alpha_{A^{}}(t)-1}
\nonumber \\ && \mbox{} \times
\frac{\pi{\alpha^\prime}_{A^{}}}{\sin[\pi\alpha_{A^{}}(t)]}
\frac{\sigma_{A^{}}+e^{-i\pi\alpha_{A^{}}(t)}}{2\Gamma[\alpha_{A^{}}(t)]}.
\label{prop}
\end{eqnarray}
Here, $s \approx 2p \cdot q = 2M_N \nu$, $t=(p-q)^2$,
and $s_0\approx 4$ GeV$^2$~\cite{DL00,KMOV00}.
The signature of the $a_1^{}$ trajectory is $\sigma_A^{}=-1$
and the Regge trajectory of the $a_1^{}$-meson is given by
\begin{equation}
\alpha_A^{}(t) = \alpha_A^{}(0) + \alpha^\prime_A t.
\label{traj}
\end{equation}
We assume that the $a_1^{}$-trajectory is an ordinary Regge
trajectory with slope $\alpha^\prime_A \approx
0.9$~GeV$^{-2}$. Then the intercept of the trajectory is estimated
as $\alpha_A^{}(0)\approx -0.36$ for the $a_1^{}$-meson mass
$M_A = 1.23$~GeV~\cite{PDG}.

\begin{figure*}[t]
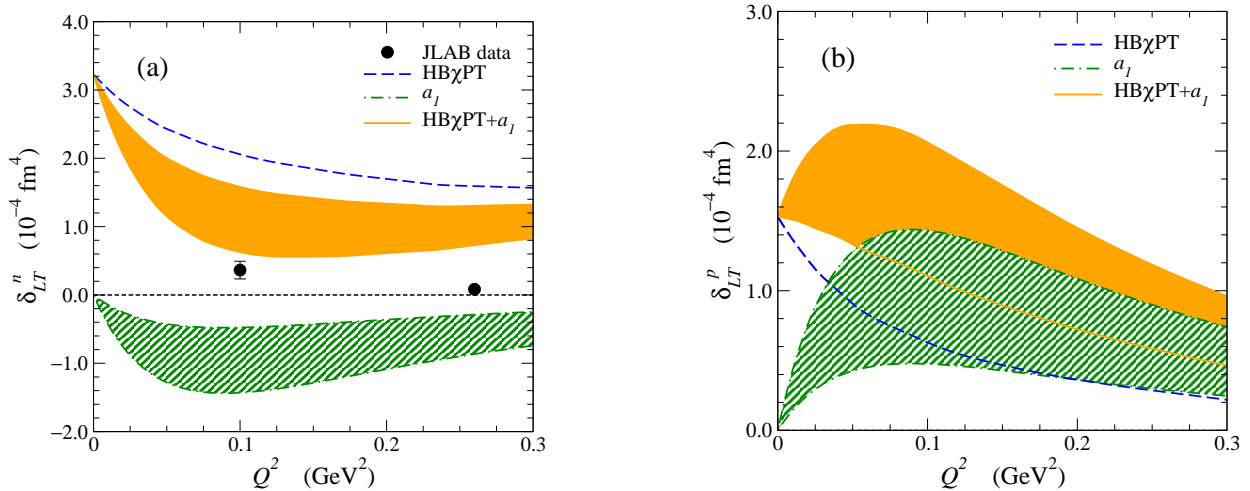
\centering
\vskip -0.4cm
\includegraphics[width=0.4\textwidth,angle=0,clip]{fig2a.eps}
\qquad\qquad\qquad
\includegraphics[width=0.4\textwidth,angle=0,clip]{fig2b.eps}
\caption{\label{fig:results}
(a) Contribution of the $a_1^{}$-exchange to the generalized
longitudinal-transverse polarizability of the neutron $\delta_{LT}^n$
as a function of $Q^2$ (shaded area). The dashed line is the result of
heavy baryon $\chi$PT of Ref.~\cite{KSV02}. The filled area is the sum of the
prediction of Ref.~\cite{KSV02} and the $a_1^{}$-exchange contribution
obtained in this work. The experimental data are from Ref.~\cite{JLE94010-04}.
(b) Same for the generalized longitudinal-transverse polarizability
of the proton $\delta_{LT}^p$.}
\end{figure*}

By making use of the relation,
\begin{equation}
\bar N(p,S) \gamma_\beta^{} \gamma_5^{} N(p,S) = 2 S_\beta,
\end{equation}
the contribution of the $a_1^{}$-trajectory exchange to $g_1^{} + g_2^{}$ is
obtained as
\begin{eqnarray}
g_1^{}(\nu,Q^2) + g_2^{}(\nu,Q^2) &=&
-g_{a_1^{}qq}^{} g_{a_1^{}NN}^{} \mathcal{A} \frac{M_N\nu \alpha^\prime_A}
{4\pi^2\Gamma[\alpha_A^{}(t)]}
\nonumber \\ && \hspace{-2cm} \mbox{} \times
\left( \frac{2M_N\nu}{s_0^{}} \right)^{\alpha_A^{}(0)-1}
F(Q^2,m_q^2).
\label{g12}
\end{eqnarray}
Substituting Eq.~(\ref{g12}) into Eq.~(\ref{def1}) and then performing
the integration over $\nu$ leads to the final result for the
$a_1^{}$ contribution to $\delta_{LT}^{}$, which reads
\begin{eqnarray}
\delta^{n,p}_{LT} &=&
\pm \frac{3 \alpha g^2_{a_1^{}pp} \alpha^\prime_A 2^{\alpha_A^{}(0)-2.5}}
{5 \pi^2 \Gamma[\alpha_A^{}(0)] \{3-\alpha_A^{}(0) \}
M_N^2 z_0^{3-\alpha_A^{}(0)} }
\nonumber \\ && \mbox{} \times
\left( \frac{M_N^2}{s_0^{}} \right)^{\alpha_A^{}(0)-1} F(Q^2,m_q^2),
\label{final2}
\end{eqnarray}
where $z_0^{} = \nu_0^{}/M_N$, and we used the constituent quark model relation~\cite{BW75}
\begin{equation}
g_{a_1^{}qq}^{} = \frac{3 g_{a_1^{}pp}^{}}{5}.
\label{constituent}
\end{equation}
The upper and the lower sign in Eq.~(\ref{final2}) correspond to the
neutron and the proton case, respectively.
In Fig.~\ref{fig:results}, our results for the $a_1^{}$-exchange contribution
to the generalized longitudinal-transverse polarizability $\delta_{LT}^{}$
of the nucleon are presented with the coupling constant
\begin{equation}
g_{a_1^{}pp}^{} = 7.15\pm 1.92
\label{valuecoupling}
\end{equation}
that is obtained by using the assumption of axial-vector dominance~\cite{BF96}, 
which gives the relation
\begin{equation}
\frac{g_A}{g_V}=\frac{\sqrt{2}f_{a_1^{}}g_{a_1^{}NN}}{m^2_{a_1^{}}},
\label{coupling}
\end{equation}
with $g_A/g_V=1.2694 \pm 0028 $ and $f_{a_1^{}}=(0.19 \pm 0.03)$ GeV$^2$~\cite{PDG}.
This value is close to the estimation of $g_{a_1^{}pp}^{}=6.13 \sim 7.09$ that was
obtained from the nucleon-nucleon potential in Ref.~\cite{Durso:1984um}.

Because we are using the constituent quark model relation in Eq.~(\ref{constituent}),
we use the constituent quark mass in the form factor of Eq.~(\ref{form}) for consistency.
In the present work, we use $m_q^{}=0.27$~GeV that is supported by the study 
on hadron spectroscopy within Dyson-Schwinger equation approach~\cite{CRT11}.%
\footnote{As the quark mass in the triangle graph becomes smaller, the contribution from
the $a_1^{}$ trajectory increases. For example, if we use $m_q^{}=170$~MeV as given by 
the mean field approximation in the instanton liquid model of Shuryak~\cite{shuryak}, 
it leads to the enhancement factor of $1.6 \sim 1.9$ to the form factor in the interval of
$Q^2=0.1-0.26$ GeV$^2$. 
Such small value of constituent quark mass is supported, for example, by the NLO calculation 
of the hadronic contribution to muon anomalous magnetic moment~\cite{Pivovarov:2001mw}. 
In principle, one may try to use the running quark mass as a function
of the quark virtuality $k^2$ in the triangle diagram as 
$m(k^2)=m_{\rm current}^{} + m_q^{}(k^2)$. 
But this causes an additional unknown form factor in the $a_1^{}$-quark vertex. 
 }

Shown in Fig.~\ref{fig:results}(a) are the results for the neutron $\delta_{LT}^{n}$,
while those for the proton $\delta_{LT}^{p}$ are given in Fig.~\ref{fig:results}(b).
Here, the dashed lines are the predictions of the heavy baryon $\chi$PT
of Ref.~\cite{KSV02}, which evidently overestimates the experimental data
for $\delta_{LT}^{n}$.
The contributions from the $a_1^{}$ exchange in Eq.~(\ref{final2}) is given by the
shaded areas because of the uncertainty of the coupling constants.
As can be seen in Fig.~\ref{fig:results}(a), the
$a_1^{}$-exchange  contribution to $\delta_{LT}^{n}$ is large and negative,
so, when combined with the $\chi$PT calculation, it can bring down
the theoretical prediction for $\delta_{LT}^{n}$ closer to the
measured data of the Jefferson Lab E94010 Collaboration~\cite{JLE94010-04}
than the $\chi$PT calculation.
By the filled area in Fig.~\ref{fig:results}(a), we give the result
that is obtained by combining the $\chi$PT prediction of Ref.~\cite{KSV02}
and the present work. This evidently shows the nontrivial role of the axial
anomaly in $\delta_{LT}^{}$.

In the case of the proton, the $a_1^{}$-exchange contribution
to $\delta_{LT}^{p}$ has the opposite sign compared to the neutron case
due to the isovector nature of the $a_1^{}$-meson as can be seen
in Fig.~\ref{fig:results}(b).
As a consequence, the contribution of the $a_1^{}$-exchange is added on
the $\chi$PT prediction and the final result becomes larger.
This is shown explicitly by the filled area in Fig.~\ref{fig:results}(b).
Thus, measuring $\delta_{LT}^{p}$, which is planned at the TJNAF~\cite{Slifer09},
is an ideal tool to test the role of the axial anomaly to $\delta_{LT}^{}$.

\begin{figure*}[t]
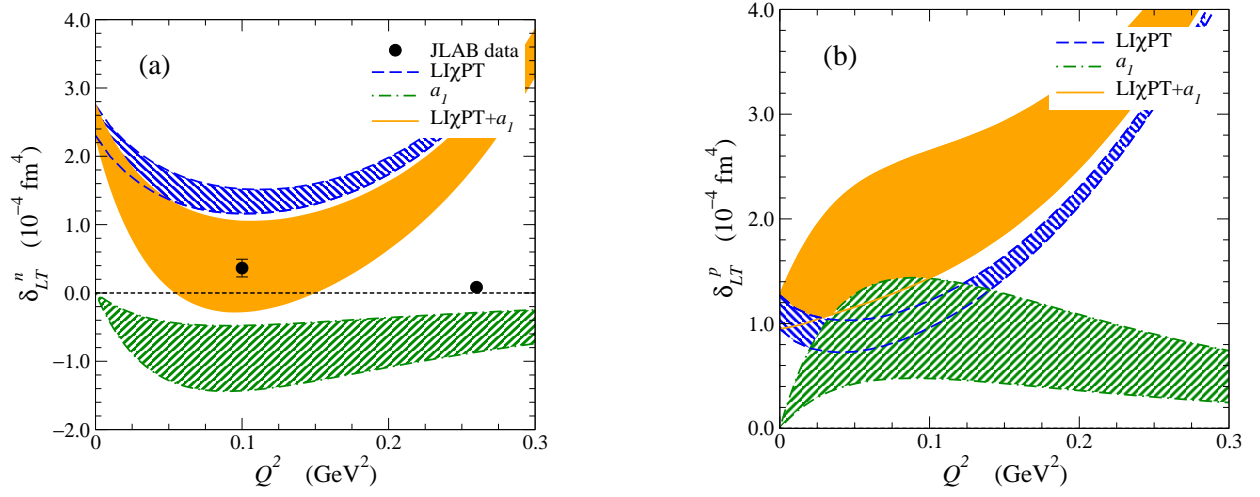
\centering
\vskip -0.4cm
\includegraphics[width=0.4\textwidth,angle=0,clip]{fig3a.eps}
\qquad\qquad\qquad
\includegraphics[width=0.4\textwidth,angle=0,clip]{fig3b.eps}
\caption{\label{fig:results2}
(a) The generalized longitudinal-transverse polarizability
of the neutron $\delta_{LT}^{n}$ and (b) of the proton $\delta_{LT}^{p}$.
Same as Fig.~\ref{fig:results} but with the result of
Lorentz-invariant $\chi$PT of Ref.~\cite{BHM03}.
}
\end{figure*}

However, different approaches of $\chi$PT give different predictions for
$\delta_{LT}^{}$ at low $Q^2$.
As can be seen in Ref.~\cite{Chen10}, the predicted $\delta_{LT}^{n}$ of
Ref.~\cite{BHM03}, which is based on a Lorentz-invariant formulation of
$\chi$PT, has a very different $Q^2$ dependence.
Namely, the predicted $\delta_{LT}^{n}$ and $\delta_{LT}^p$
of Ref.~\cite{BHM03} decrease with $Q^2$ at low $Q^2$ region and then
increase when $Q^2 \ge 0.05 \sim 0.1$~GeV$^2$, while those of
Ref.~\cite{KSV02} monotonically decrease with $Q^2$.
This shows that $\delta_{LT}^{}$
is very sensitive to the approach to hadron reactions.
In Fig.~\ref{fig:results2}, we give the results as in
Fig.~\ref{fig:results} but with the calculation of Ref.~\cite{BHM03}.
The results for $\delta_{LT}^{n}$ are shown in Fig.~\ref{fig:results2}(a),
while those for $\delta_{LT}^{p}$ are in Fig.~\ref{fig:results2}(b).
Since the contribution from the $\Delta$ resonance is hard to control,
the predictions of Lorentz-invariant $\chi$PT are given by the
shaded areas with dashed lines in Fig.~\ref{fig:results2} following Ref.~\cite{BHM03}.
This evidently shows that, although the two approaches of $\chi$PT give very different
predictions on $\delta_{LT}^{}$, both of them overestimate the measured
$\delta_{LT}^{}$ of the neutron.
Combining the $a_1^{}$-exchange with the prediction of Ref.~\cite{BHM03} leads
again to a better agreement with the measured data for $\delta_{LT}^{n}$.
Thus, as can be seen in Figs.~\ref{fig:results}(a) and \ref{fig:results2}(a), the
$a_1^{}$-exchange has a crucial role to bring the $\chi$PT calculations of
Refs.~\cite{KSV02,BHM03} to the measured data for $\delta_{LT}^{n}$, in particular, at
low $Q^2$ region.
However, we still overestimate the data at $Q^2 = 0.26$~GeV$^2$ for the both cases,
and more elaborated examinations on the $Q^2 > 0.2$~GeV$^2$ region
as well as on the $\chi$PT approaches are awaited.

Finally, we make a comment on the contribution of the flavor singlet
$f_1(1285)$ axial-vector meson exchange to $\delta_{LT}^{}$.
Because of the small coupling constant $g_{f_1^{}NN}^{} \sim 2.5$~\cite{BF96}
and the constituent quark model relation
$g_{f_1^{}qq}^{} = g_{f_1^{}NN}^{}/3$, the contribution
of the $f_1$-exchange is found to be an order of magnitude smaller than that of
the $a_1^{}$-exchange. This allows us to safely ignore
the $f_1$-exchange contribution in the considered kinematical region.

In summary, we calculated the contribution of the axial anomaly to the
generalized longitudinal-transverse polarizability $\delta_{LT}^{}$
through the exchange of the $a_1^{}$ Regge trajectory. In spite of the large
uncertainties in the $a_1^{}$ trajectory exchange, we found that
its contribution is nontrivial and large enough to
counterbalance the discrepancy between the $\chi$PT predictions
and the experimental data for the neutron, especially, at low $Q^2$ region.
To further test the role of the axial anomaly in the
generalized longitudinal-transverse polarizability, we also presented
the prediction for $\delta_{LT}$ of the proton, which will be measured at
TJNAF in near future.

\acknowledgments

We are very grateful to A.~E. Dorokhov and S.~B. Gerasimov for
fruitful discussions.
N.K. acknowledges the warm hospitality of the Department of Physics of
Kyungpook National University where a part of this work was done.
The work of N.K. was supported in part by the RFBR Grant 10-02-00368-a
and the JINR-Belarus N183 project.
Y.O. was supported by the Basic Science Research Program through the
National Research Foundation of Korea (NRF) funded by the Ministry
of Education, Science and Technology (Grant \mbox{No.} 2010-0009381).


\begin{thebibliography}{10}

\bibitem{KSV02}
C.~W. Kao, T.~Spitzenberg, and M.~Vanderhaeghen,
\newblock Phys. Rev. D \textbf{67}, 016001 (2003).

\bibitem{BHM03}
V.~Bernard, T.~R. Hemmert, and U.-G. Mei{\ss}ner,
\newblock Phys. Rev. D \textbf{67}, 076008 (2003).

\bibitem{JLE94010-04}
Jefferson Lab E94010 Collaboration, M.~Amarian \textit{et~al.\/},
\newblock Phys. Rev. Lett. \textbf{93}, 152301 (2004).

\bibitem{Slifer09}
K.~Slifer,
\newblock AIP Conf. Proc. \textbf{1155}, 125 (2009).

\bibitem{Adler69}
S.~L. Adler,
\newblock Phys. Rev. \textbf{177}, 2426 (1969).

\bibitem{BJ69}
J.~S. Bell and R.~Jackiw,
\newblock Nuovo Cimento \textbf{60}, 47 (1969).

\bibitem{AEL95}
See, for example, M.~Anselmino, A.~Efremov, and E.~Leader,
\newblock Phys. Rep. \textbf{261}, 1 (1995).

\bibitem{BK96d}
J.~Blumlein and N.~Kochelev,
\newblock Nucl. Phys. B \textbf{498}, 285 (1997).

\bibitem{BKRS99}
J.~C.~R. Bloch, \mbox{Yu}.~L. Kalinovsky, C.~D. Roberts, and S.~M. Schmidt,
\newblock Phys. Rev. D \textbf{60}, 111502(R) (1999).

\bibitem{Landau48}
L.~D. Landau,
\newblock Dokl. Akad. Nauk USSR \textbf{60}, 207 (1948).

\bibitem{Yang50}
C.~N. Yang,
\newblock Phys. Rev. \textbf{77}, 242 (1950).

\bibitem{GLV97}
M.~Guidal, J.-M. Laget, and M.~Vanderhaeghen,
\newblock Nucl. Phys. A \textbf{627}, 645 (1997).

\bibitem{DL00}
A.~Donnachie and P.~V. Landshoff,
\newblock Phys. Lett. B \textbf{478}, 146 (2000).

\bibitem{KMOV00}
N.~I. Kochelev \textit{et~al.\/},
\newblock Phys. Rev. D \textbf{61}, 094008 (2000).

\bibitem{PDG}
\newblock Particle Data Group, K. Nakamura \textit{et~al.\/}, 
J. Phys. G \textbf{37}, 075021 (2010).

\bibitem{BW75}
G.~E. Brown and W.~Weise,
\newblock Phys. Rep. \textbf{22}, 279 (1975).

\bibitem{BF96}
M.~Birkel and H.~Fritzsch,
\newblock Phys. Rev. D \textbf{53}, 6195 (1996).

\bibitem{Durso:1984um}
J.~W.~Durso, G.~E.~Brown, and M.~Saarela,
\newblock  Nucl. Phys. A  {\bf 430}, 653 (1984).

\bibitem{CRT11}
See, for example, L.~Chang, C.~D. Roberts, and P.~C. Tandy,
\newblock arXiv:1107.4003.

\bibitem{shuryak} 
T. Sch\"afer and E.~V. Shuryak,
\newblock Rev. Mod. Phys. {\bf 70}, 323 (1998).

\bibitem{Pivovarov:2001mw}
A.~A.~Pivovarov,
\newblock Yad. Fiz. \textbf{66}, 934 (2003) [Phys. Atom. Nucl.  \textbf{66}, 902 (2003)].

\bibitem{Chen10}
J.-P. Chen,
\newblock Int. J. Mod. Phys. E \textbf{19}, 1893 (2010).



\end{thebibliography}
\end{document}